# Atomic-ordering-induced quantum phase transition between topological crystalline insulator and $Z_2$ topological insulator

Hui-Xiong Deng[1*], Zhi-Gang Song[1], Shu-Shen Li[1,2], Su-Huai Wei[3*] and Jun-Wei Luo[1,2*]

[1]*State Key Laboratory of Superlattices and Microstructures, Institute of Semiconductors, Chinese Academy of Sciences, Beijing 100083, China*

[2]*Synergetic Innovation Center of Quantum Information and Quantum Physics, University of Science and Technology of China, Hefei, Anhui 230026, China*

[3]*Beijing Computational Science Research Center, Beijing 100094, China*

e-mail address: hxdeng@semi.ac.cn

e-mail address: suhuaiwei@csrc.ac.cn

e-mail address: jwluo@semi.ac.cn

**Topological phase transition in a single material usually refers to transitions between a trivial band insulator and a topological Dirac phase, but the transition may also occur between different classes of topological Dirac phases. However, it is a fundamental challenge to realize quantum transition between $Z_2$ nontrivial topological insulator (TI) and topological crystalline insulator (TCI) in one material because $Z_2$ TI and TCI are hardly both co–exist in a single material due to their contradictory requirement on the number of band inversions. The $Z_2$ TIs must have an odd number of band inversions over all the time-reversal invariant momenta, whereas, the newly discovered TCIs, as a distinct class of the topological Dirac materials protected by the underlying crystalline symmetry, owns an even number of band inversions. Here, take PbSnTe$_2$ alloy as an example, we show that at proper alloy composition the atomic-ordering is an effective way to tune the symmetry of the alloy so that we can electrically switch between TCI phase and $Z_2$ TI phase when the alloy is ordered from a random phase into a stable CuPt phase. Our results suggest that atomic-ordering provides a new platform to switch between different topological phases.**

Topological insulators (TIs) [1, 2, 3, 4, 5, 6, 7, 8] are an emerging class of quantum materials, which are nontrivial under the $Z_2$ topological classification (i.e., $Z_2$ is an odd number) usually resulting from band inversions occurring at an odd number of time-reversal-invariant momenta (TRIMs). These materials possess topological surface states spanning the insulating bulk bandgap, when they are placed next to a vacuum or a $Z_2$ topological trivial material, owing to the impossible change of the characterized topological invariant in crossing the interface between them without closing the band gap [9]. These spin-momentum-locked helical surface states exhibit Dirac-cone energy dispersion across the bulk bandgap, and are topologically protected by the time-reversal symmetry (TRS) [2,6,7]. Such topological band insulators may provide new routes for generating novel phases and particles, such as hosting Majorana quasiparticles as well as the condensed-matter realization of magnetic monopole-like behavior, possibly finding potential application in spintronics and quantum computing[9].

Some materials such as SnTe also possess band inversions, but they are $Z_2$ topological trivial because their band inversions occur at an even number of TRIMs (e.g., SnTe band inversion occur at 4 $L$-points)[10, 11]. Recent theoretical results followed by experimental validations have suggested that these materials are topological crystalline insulators (TCIs) [12, 13, 14, 15], a subclass of topological insulators in which the underlying crystalline symmetry replaces the role of TRS in ensuring a new topological invariant of mirror Chern number instead of the $Z_2$ index.

Despite their common characteristics of topological protected gapless spin-momentum-locked surface states and an intrinsic orbital texture switch occurring exactly at the Dirac point [16, 17], $Z_2$ TI and TCI have the very distinct topology of surface electronic structures[18, 19, 20]. For instance, the Dirac points in $Z_2$ TIs are nailed to TRIMs as a result of TRS protection[4, 5, 6, 7, 8, 21], whereas, the Dirac points in TCIs are not pinned at the TRIMs[13, 14, 15, 18, 19, 20], demonstrating their irrelevance to the TRS-related protection. As a consequence of different topological invariants, compared with $Z_2$ TIs, in which the topological states are robust against general time-reversal invariant perturbations, TCI surface states have a much wider range of tunable electronic properties under various perturbations, such as structural distortion, magnetic dopant, mechanical strain, thickness engineering, and disorder. Topological surface states in the $Z_2$ TIs are susceptible to TRS-breaking disorders such as magnetic defects; however, it is also possible to realize magnetic yet topologically protected surface states in the TCI system due to its irrelevance to the TRS, which is fundamentally distinct from the $Z_2$ TIs. Therefore, the magnetic and superconducting orders in the surface states in TCI can be different from those observed in the $Z_2$ TIs. Perturbations without breaking TRS in the TCIs can move Dirac points in momentum space, mimicking the effect of a gauge field vector potential[22], and open an energy gap at the Dirac point, generating Dirac mass[23, 24]. Changing the alloying composition [25] or applying a strain[20, 22] are two recently demonstrated effective ways of moving surface Dirac points in TCIs. Moreover, accompanying the formation of offspring Dirac cones, both Lifshitz transitions and Van Hove singularity (VHS) were aware of the existing in TCI band structures[17, 23]. Considering the broad applications of magnetic materials in modern electronics, such topologically protected surface states compatible with magnetism will be of considerable interest regarding integrating TI materials into future electronic devices. Therefore, the feasibility of reversible topological phase transition between $Z_2$ TI and TCI phases in a material is highly desirable to explore novel applications.

Although some attempts have been made to achieve such topological phase transition by reducing the crystal lattice symmetry through stain engineering or forming surfaces[10, 11], to the best of our knowledge, it has not been realized either in theoretical predictions or experimental observations. For instance, Fu et. al. [10], proposed in their pioneering study of $Z_2$ TIs that, an uniaxial strain applied along the [111] direction to the $Pb_xSn_{1-x}Te$ alloy separates in energy the $L$ point along the [111] is separated from remaining three $L$ points and may result in an odd number of band inversions (thus realizing the $Z_2$ TI phase) at some composition $x$. However, we found such proposal will not work because $Pb_xSn_{1-x}Te$ alloy becomes metallic before occurring the odd number of band inversions, after analyzing the band structure of PbTe and SnTe (Figure S1 in the Supplemental Material). Here, we alternatively propose to achieve the quantum phase transition between $Z_2$ TI and TCI via atomic-ordering. Taking $PbSnTe_2$ alloy as the prototype, we show that the atomic-ordering of this alloy into the CuPt phase breaks the four (parent) equivalent rocksalt $L$-points into one (child) $\bar{\Gamma}$ and three (child) $\bar{F}$ points with non-identical bandgaps. Subsequently an electrical controllable strain applied to the CuPt-ordered PbSnTe2 alloy drives gap-closing first at child $\bar{\Gamma}$-point (becoming odd number of band inversions) and then at $\bar{F}$ -point (becoming a normal insulator), as schematically shown in Fig. 1a, consequently, achieving a reversible quantum transition between $Z_2$ TI and TCI phases.

The group IV chalcogenide SnTe is the prototype TCI. It possesses a rocksalt crystal structure with $O_h$ symmetry at room temperature, and its fundamental band gaps occur at four equivalent $L$ points in the face-centered-cubic (FCC) Brillouin zone (BZ). Despite the fact that Sn sits between Ge and Pb in the same column of the periodic table, SnTe is a TCI with inverted bandgaps but PbTe and GeTe are normal trivial insulators [27]. We can easily engineer these (rocksalt) group IV chalcogenides from normal insulator to TCI or vice versa by forming random mixed-cation chalcogenide alloys[28], but not from TCI to $Z_2$ TI because the bandgaps of these alloys always occur at even (four) equivalent $L$ points. However, the quantum phase transition from TCI to $Z_2$ TI may become possible if the alloys form lower symmetry ordered phases that break the symmetry and the equivalence of $L$ points in rocksalt structure. To see whether such ordered phases can form at low temperature, we have calculated, using the first-principles density functional theory (DFT), the alloy formation energies $\Delta H_f(\sigma, A_{1-x}B_xC) = E(\sigma, A_{1-x}B_xC) - [(1-x)E(AC) + xE(BC)]$ of the group-IV telluride and selenide alloys at $x$=0.5 (with configuration $\sigma$ in random structure and common ordered alloy structures[29]). The ordered configurations $\sigma$ we studied are CuPt [(1,1) superlattice along the (111) direction], CuAu [(1,1) superlattice along the (001) direction], chalcopyrite [(2,2) superlattice along the (201) direction], Y2 [(2,2) superlattice along the (110) direction], and Z2 [(2,2) superlattice along the (001) direction]. The random alloy is modeled here using the "special quasi random structures" (SQS) approach[30]. These results are summarized in Table SI in the Supplemental Material. We find that the CuPt-ordered structure with a point group of $D_{3d}$ has the lowest formation energy, indicating that rocksalt $ABC_2$ (A or B=Ge, Sn, Pb, C=Te, Se) alloys can spontaneously order in the CuPt structure, at least at low temperature during lattice matched coherent growth. This finding is expected because the CuPt structure possesses the smallest strain energy over all the alloy structures because it can allow all nearest cation-anion bonds to attain their respective ideal equilibrium lengths with the minimum bond bending [29]. As a result, the energy differences between the ground state CuPt structure and the SQS structure of the $ABC_2$ (A or B=Ge, Sn, Pb; C=Te, Se) alloys increase monotonically with the magnitude of the lattice mismatch between the two end constituents, as shown in Table SI.

The CuPt-ordered alloy structure possesses a double sized rhombohedral unit cell and thus half BZ compared to the parent rocksalt structure, along with a symmetry reduction from $O_h$ to $D_{3d}$ ($R\bar{3}m$ space group), as shown in Figs. 2a, b. As a consequence, four equivalent $L$ points in the rocksalt BZ transform separately to one $\bar{\Gamma}$ point and three $\bar{F}$ points in reduced CuPt-like structure BZ, in which $\bar{\Gamma}$ and $\bar{F}$ are no longer symmetry-equivalent [31]. The bandgaps at $\bar{\Gamma}$ and $\bar{F}$ points now become non-identical, resulting from their differences of symmetry controlled inter-band coupling. Therefore, CuPt-ordered $ABC_2$ (A or B=Ge, Sn, Pb; C=Te, Se) alloys provide an ideal test-bed for studying topological phase transition between $Z_2$ TI and TCI through engineering the band structure. If band inversion occurs only at $\bar{\Gamma}$ point but not at remaining TRIM $k$ points, the CuPt-ordered alloy becomes a $\Gamma$-phase $Z_2$ TI [26]. If band inversions occur exclusively at three equivalent $\bar{F}$ points, the alloy becomes translationally active phase TI[26]. And if like in SnTe band inversions happen at both $\bar{\Gamma}$ and $\bar{F}$ point, the alloy belongs to TCI. Figure S2 in the Supplemental Material shows the first-principles calculated band structure of CuPt ordered PbSnTe$_2$ alloy with the equilibrium lattice constant using the modified Becke and Johnson exchange potential[32]. The bandgaps at $\bar{\Gamma}$ and $\bar{F}$ points are indeed non-identical with values of 42.5 and 19.4 meV, respectively, despite both of them are derived from the symmetry equivalent $L$ points in the parent rocksalt group-IV tellurides. The energy difference of the band edge states at $\bar{\Gamma}$ and $\bar{F}$ points is mainly caused by the distinct inter-band coupling at these two symmetry points as a result of symmetry reduction from (rocksalt) $O_h$ to (CuPt-type alloy) $D_{3d}$. As shown in Fig. S2, at the equilibrium condition, the bandgap at the $\bar{F}$ point is smaller than that at the $\bar{\Gamma}$ point. By analyzing the wave function characters, we find that the conduction band edge at the $\bar{\Gamma}$-point mainly comes from the anion $p$ orbital with an even parity ("+") and the valence band edge from the cation $p$ orbital with an odd parity ("-"). Such band order is same as the $\bar{F}$-point. Consequently, both bandgaps at $\bar{\Gamma}$ and $\bar{F}$ points are inverted, indicating that the PbSnTe$_2$ alloy is a TCI [19]. Fortunately, the distinct bandgaps at $\bar{\Gamma}$ and $\bar{F}$ points, which is an unique property of CuPt ordered PbSnTe$_2$ alloy, enable us to realize the quantum phase transition from TCI to $Z_2$ TI since we could further engineer the band structure of the alloy to ensure band inversions occurring separately at $\bar{\Gamma}$ and $\bar{F}$ points, e.g., by applying strain or pressure.

The strain has been demonstrated to be a particular compelling tuning "knob" to engineer the electronic band structure[20, 22, 33], as it can tune the interatomic lattice spacing and induce an accompanying adjustment in the electronic band structure for a fixed chemical composition. In TCIs, for example, strain has been explored to generate pseudo-magnetic fields and helical flat bands [22], to engineer the phase transition from normal insulator to TCI [33], and to finely modify the characteristics of the topological surface bands [20]. Controllable manipulation of strain is necessary for the creation of suitable platforms for applications in the growing field of straintronics[20]. A promising pathway may involve the use of electric-field-induced strains [34], in which the change of inter-atomic lattice spacing is relying on the piezoelectric response of a piezoelectric material substrate to the electric field. The piezoelectric effect is the linear electromechanical interaction between the mechanical and the electrical state in a crystal. A reversible electric-field-induced strain of over 5% has been reported, e.g., in BiFeO$_3$ films[34] and piezoelectric-induced strains have been employed, e.g., to tune semiconductor quantum dots for strain-tunable entangled-light-emitting-diodes [35]. Here we propose to place the PbSnTe$_2$ alloy film onto a piezoelectric actuator via gold-thermocompression bonding, as described in Ref. [35], allowing the *in situ* application of on-demand biaxial strains by tuning the voltage (electric field), as schematically shown in Fig. 1. By sweeping the gate voltage applied to the piezoelectric actuator from negative to positive and vice versa, we can reversibly tune the biaxial strain applied to the PbSnTe$_2$ alloy from the compressive (negative) to the tensile (positive).

Upon application of 1.13% tensile strain, we indeed find that the TCI PbSnTe$_2$ alloy becomes a $Z_2$ TI, achieving a novel quantum phase transition between TCI and $Z_2$ TI. Figure 2c shows that, at the $\bar{\Gamma}$ point, the conduction band edge state mainly comes from the anion $p$ orbital with an even parity ("+") and the valence band edge state from the cation Pb $p$ orbital with an odd parity ("-"). Such band order is same as in the TCI SnTe and is thus inverted. Whereas, at $\bar{F}$ points, the conduction band edge state mainly arises from the cation $p$ orbital ("-"), and the valence band edge state from the anion $p$ orbital ("+"), being opposite to the band order in the TCI SnTe. We assign the PbSnTe$_2$ alloy under 1.13% tensile strain as a $\Gamma$-phase $Z_2$ TI regarding it has one band inversion (at the $\bar{\Gamma}$ point). To further confirm our assignment, we have calculated the $Z_2$ topological invariants (see Supplemental Materials) following the procedure introduced by Soluyanov and Vanderbilt [36]. We find its $Z_2$ being (1;000), which is an index of a strong $Z_2$ TI [26], despite that its end constitute SnTe is a TCI and PbTe is a trivial band insulator.

We next examine topological properties of PbSnTe$_2$ alloy in a wide range of strains. Figure 1b shows the band-edge evolutions of PbSnTe$_2$ alloy as a function of biaxial strain from 0.9% to 1.3%. Conduction band edges at both $\bar{\Gamma}$ and $\bar{F}$ points are shifted at the same rate to lower energy, whereas, valence band edges to higher energy at also the same rate, as reducing the strain. Such responses of conduction and valence band states to applied strain are expected for bonding and anti-bonding states, respectively. Because of distinct bandgaps at $\bar{\Gamma}$ and $\bar{F}$ points, there are two critical points $c_{\bar{F}}$ (=1.02%) and $c_{\bar{\Gamma}}$ (=1.16%), corresponding to the strains where band order changing occurs at $\bar{F}$ and $\bar{\Gamma}$ points, respectively. From band edge evolutions, we can straightforwardly find that, when strain is smaller than $c_{\bar{F}}$ (left area in Fig. 1b), band inversions occur at both $\bar{\Gamma}$ and $\bar{F}$ points in the PbSeTe$_2$ alloy, which remains a TCI being $Z_2$ trivial (i.e., $Z_2$ = (0;000)) but Mirror Chern number nontrivial. As we continuously increase the applied tensile strain to exceed $c_{\bar{F}}$ but smaller than $c_{\bar{\Gamma}}$ (middle area in Fig. 1b), band order changes to normal at $\bar{F}$ points but remains inverted at the $\bar{\Gamma}$ point, therefore, PbSeTe$_2$ alloy becomes a $Z_2$ TI (i.e., $Z_2$ = (1;000), as shown in the bottom of Fig. 1b). When we further increase the applied strain larger than $c_{\bar{\Gamma}}$, there is no band inversion occurring in PbSeTe$_2$ alloy at both $\bar{\Gamma}$ and $\bar{F}$ points or any other $k$ points, indicating PbSeTe$_2$ alloy within this strain range being a trivial normal insulator (i.e., $Z_2$ = (0;000)). Therefore, we have demonstrated, for the first time, that the novel quantum phase transition between TCI and $Z_2$ TI, in addition to the phase transition between normal insulator and $Z_2$ TI, can be realized by atomic ordering and strain, e.g., controlled by the gate electric fields via the piezoelectric effect. These transitions can be done reversibly using the electric field to control the strain, as schematically illustrated in Figure 1a.

It is interesting to note that the strain can also drive the PbSnTe$_2$ alloy to become a three-dimensional (3D) topological Dirac materials (TDMs) [37, 38, 39], at two transition points $c_{\bar{\Gamma}}$ (between $Z_2$ TI and normal insulator) and $c_{\bar{F}}$ (between $Z_2$ TI and TCI), respectively. 3D TDMs are another novel state of quantum matter, being viewed as a 3D graphene with linear energy dispersions along all three-dimensional momentum directions instead of in a two-dimensional (2D) plane of the Dirac fermions in graphene or on the surface of 3D topological insulators. At transition point $c_{\bar{\Gamma}}$, PbSnTe$_2$ alloy owns a single 3D Dirac cone, whereas, at $c_{\bar{F}}$ point, it possesses three 3D Dirac cones. Each 3D Dirac cone in a TDM is composed of two overlapping Weyl fermions[37], which can be separated in the momentum space, by breaking the time reversal or inversion symmetry, to form the topological Weyl semimetal, a new topological quantum state exhibiting unique Fermi arcs on the surface[40, 41].

Having realized the transition between $Z_2$ TI and TCI, we turn to compare the topological surface band structures between $Z_2$ TI and TCI phases in the CuPt-ordered PbSnTe$_2$ alloy. In the TCI phase, the nontrivial topological surface band structure with an even number of Dirac cones is expected to arise only on the surfaces containing at least one of the {110} mirror planes, in which the CuPt-ordered structure is symmetric about the {110} mirror planes. Whereas, in the $Z_2$ TI phase, the topological surface states emerge on any surfaces owing to they are protected by the TRS rather than crystalline symmetries. To verify these, we calculate band structures of [001]-oriented slabs made of CuPt-ordered PbSnTe$_2$ alloy, which is symmetric through the (110) mirror plane, in TCI phase (Fig. 3c, under zero strain) and $Z_2$ TI phase (Fig. 3d, under tensile strain of 1.07%), respectively (see Supplemental Material for more details). For [001]-oriented slabs, the $\bar{\Gamma}$ point in the CuPt-ordered alloy BZ remains at the projected zone center $\bar{\bar{\Gamma}}$ in the surface BZ, whereas, one of the $\bar{F}$ points projects to the point of (0, $b_2$), and the remaining two $\bar{F}$ points to the $\bar{\bar{X}}$-point of (1/2$b_1$, 0) (where $b_1$ and $b_2$ are the reciprocal lattice vectors of the surface BZ (Fig. 3b)). Apparently, the (0, $b_2$) point is equivalent to the $\bar{\bar{\Gamma}}$-point since they are just different by a reciprocal vector of the surface BZ. Consequently, in the TCI phase, there are two Dirac cones centered at the $\bar{\bar{\Gamma}}$-point of the surface BZ as a result of the band inversions occurring at both $\bar{\Gamma}$ and $\bar{F}$ points. Whereas in the $Z_2$ TI phase, there is only one Dirac cone centered at the $\bar{\bar{\Gamma}}$-point because the band inversion exists exclusively at the $\bar{\bar{\Gamma}}$-point. We, therefore, expect to see significant differences in the surface band structures between TCI and $Z_2$ TI phases even in the same CuPt-ordered PbSnTe$_2$ alloy.

Figure 3c shows that the predicted surface band structure of a 21 atomic layers (ALs) thick PbSnTe$_2$ alloy slab in the TCI phase is rather complex and involves multiple Dirac cones, as expected for TCIs[13].

It consists of two parent Dirac cones centered at the $\bar{\bar{\Gamma}}$-point and vertically offset in energy owing to symmetry enforced coupling between these two Dirac cones. When they come together away from the $\bar{\bar{\Gamma}}$-point, the coupling between the lower half of the upper parent Dirac cone and the upper half of the lower parent Dirac cone opens a gap at all points except along the mirror line, leading to the formation of a pair of offspring Dirac points shifted away from the $\bar{\bar{\Gamma}}$-point. In the slab surface BZ, the $\bar{\bar{\Gamma}}$-$\bar{\bar{Y}}$ line is a mirror line, which is along the {110} reflection axis. The mirror Chern number is invariant under reflection about the {110} reflection axis, except for the k-points on the $\bar{\bar{\Gamma}}$-$\bar{\bar{Y}}$ line (Fig. 3b). Thus, all of the k-points off the $\bar{\bar{\Gamma}}$-$\bar{\bar{Y}}$ line have the same mirror symmetry and present the same mirror eigenvalues, whereas the k-points on the $\bar{\bar{\Gamma}}$-$\bar{\bar{Y}}$ line hold the opposite mirror eigenvalues[13, 18]. Consequently, the interaction between the two parent Dirac cones is forbidden along the $\bar{\bar{\Gamma}}$-$\bar{\bar{Y}}$ direction and crosses each other, and such interaction is allowed along remaining k-points and opens a gap, leading to the formation of a pair of offspring Dirac cones. This surface band structure is consistent with that of the prototype TCI SnTe, which was discovered theoretically[13, 18, 22] and confirmed experimentally[38], in the vicinity of Dirac points, except for two striking distinctions. Specifically, we find here that the parent Dirac cones center at the BZ center rather than at the BZ boundary, and hence, in compared with the TCI SnTe, there is only one pair of offspring Dirac cones instead of two pairs. These two distinctions are consequences of band inversions occurring at different TRIMs in CuPt-ordered $PbSnTe_2$ alloy and SnTe, respectively. Accompanying the formation of offspring Dirac cones, more striking features could also be found in the surface band structure of the TCI phase, such as recently revealed the existence of a Lifshitz transition as a result of switching of the orbital characters of the upper parent Dirac cone and the lower parent Dirac cone[17]. The saddle point in the TCI surface band structure (along the *k*-line perpendicular to the mirror line away from the $\bar{\bar{\Gamma}}$-point, as shown in Fig. 3c) is known as the Van Hove singularity

(VHS)[23]. The existence of both Lifshitz transitions and VHSs provides the possibility of achieving future quantum applications[23].

Figure 3d shows the corresponding surface band structure in the $Z_2$ TI phase. In compared with a complex band structure involving multiple Dirac cones in the TCI phase, the surface band structure in the $Z_2$ TI phase is quite simple. In general, we expect the Dirac cones centered exactly at the $\bar{\bar{\Gamma}}$-point in the $Z_2$ TI, owing to the TRS protects the topological states, and, subsequently, without offspring Dirac cones and absence of the Lifshitz transition points. However, we observe a finite Dirac gap of about 0.08 eV as a result of strong hybridization between the two topological surface states (TSSs) located on the top and bottom surfaces, respectively. Figure 3g shows, as expected, that the magnitude of the surface bandgap decreases monotonically as the thickness of the slab increases. Such surface bandgap is an intrinsic feature of finite thick slabs and films[42], and was frequently observed in thin films of prototype 3D chalcogenide TIs[43, 44]. In order to mimic the surface states of the semi-infinite slabs, we also calculate the surface band structure based on the surface Green Function with a low-energy effective $k \cdot p$ Hamiltonian model (see the Supplemental Material for more details ). We find that there is a gapless Dirac cone at the $\bar{\bar{\Gamma}}$ point formed by the surface states in the semi-infinite slab of the PbSnTe$_2$ alloy, as shown in the Fig. 3h.

**Methods**

To engineer the band structure of alloys, we perform the first-principles calculations using the frozen-core projector-augmented wave method (PAW) within the density functional theory (DFT)[45, 46] and generalized-gradient-approximation (GGA) to the exchange-correlation potential as implemented in the VASP code[47, 48, 49]. We have carefully examined the convergence of the plane-wave cutoff energy and the Monkhorst-Pack k-point mesh. In all calculations, all the atoms are allowed to relax until the quantum mechanical forces acting on them become less than 0.02 eV/Å. The spin-orbit coupling (SOC) term is included. Regarding semi-local exchange-correlation functionals, such as GGA, usually remarkably underestimate the bandgaps of semiconductors, as a comparison, we have also employed the modified Becke and Johnson exchange potential[32] (hereafter called TB-MBJ), which has the comparable accuracy with the hybrid functional and GW method, but much less computational cost. For example, the TB-MBJ calculated normal gap of PbTe and inverted gap of SnTe are 0.20 eV and -0.19 eV, respectively, very close to experimental values. The random alloys are mimicked by the special quasirandom structures (SQS)[30, 50]. To obtain the surface band structure of the semi-infinite slabs, we calculated the surface Green Function based on the low-energy effective $k \cdot p$ Hamiltonian model. The calculation of the $Z_2$ invariant is based on the time-reversal polarization theory developed by Fu and Kane[51], in which one only needs to track the Wannier charge centers (WCCs) spectrum $\bar{\chi}_n$ and determine the center of the largest gap $z^m$ between two adjacent WCCs on the circle[36]. More details can be found in the Supplemental Information.

**Acknowledgements**

This work was supported by the National Natural Science Foundation of China (NSFC) under Grants Nos. 61121491, 11474273, 11104264 and U1530401. J. W. L. was also supported by the National Young 1000 Talents Plan.

**Additional Information**

Supplementary information is available for this paper. Reprints and permissions information is available online at www.nature.com/reprints. Correspondence and requests for materials should be addressed to J. W. Luo (jwluo@semi.ac.cn).

**Competing financial interests**

The authors declare no competing financial interests.

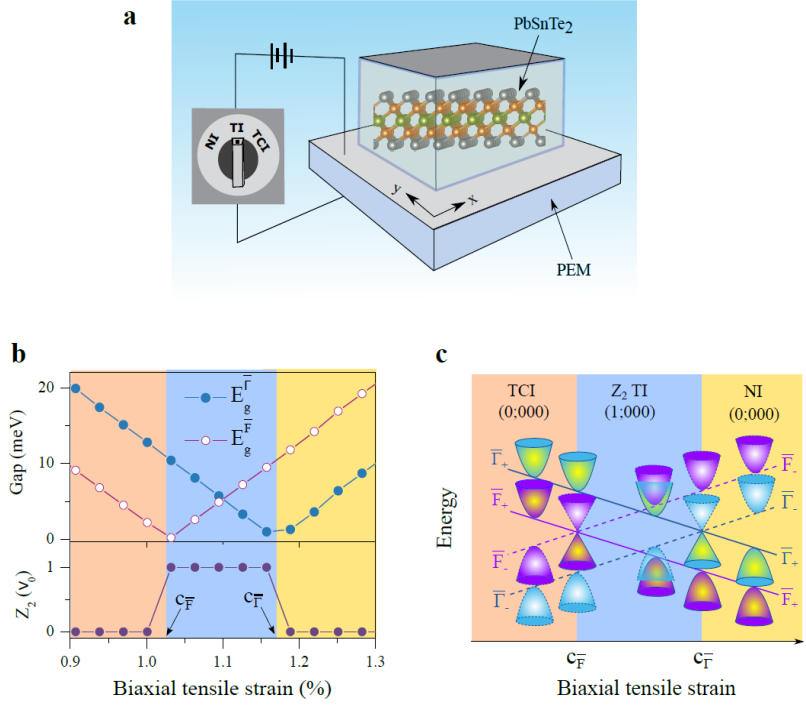

FIG. 1. (Color online) **Electrically control of reversible quantum phase transition between topological crystalline insulator (TCI) and $Z_2$ topological insulator TI. a,** Sketch of the strain-tunable topological phase transition among topological crystalline insulator (TCI), topological insulator (TI), and normal insulator (NI). The top layer is the CuPt ordered PbSnTe$_2$ alloy. The bottom layer is the piezoelectric materials (PEM), which can change the in-plane strain by applying voltage. **b,** Band gaps at $\bar{\Gamma}$ and $\bar{F}$ points as a function of biaxial strain, and the topological invariant quantity $Z_2$ (here only the strong topological index $v_0$ is shown for 3D system) as a function of biaxial strain for the CuPt ordered alloy. **c,** The schematic evolution of band-edges at $\bar{\Gamma}$ and $\bar{F}$ points as a function of biaxial strain (0.9%-1.3%) in the CuPt-ordered PbSnTe$_2$ alloy. The subscript (minus and plus signs) represent the odd and even parities, respectively.

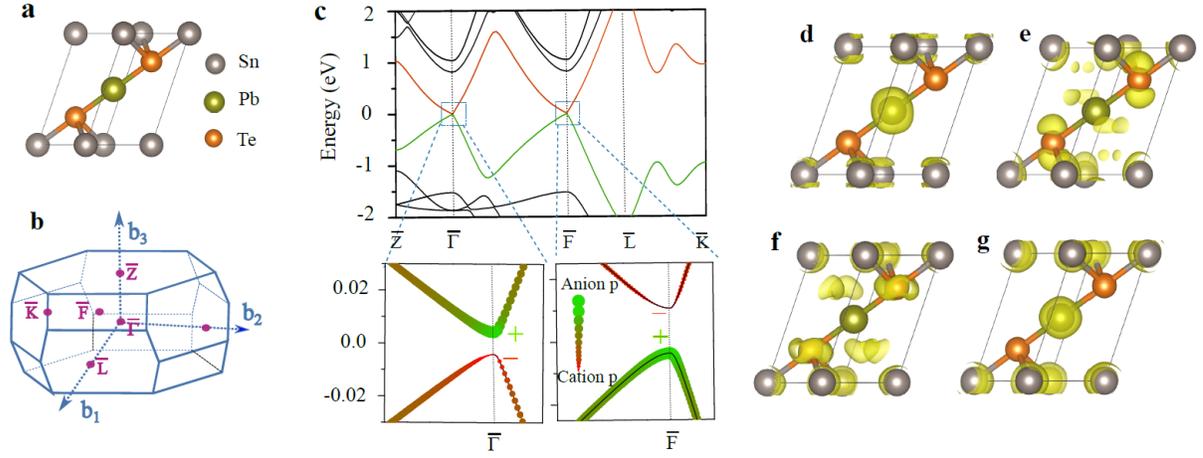

FIG. 2. (Color online) **Band structure of CuPt-order PbSnTe$_2$ alloy under biaxial strain. a**, Crystal structures of CuPt ordered PbSnTe$_2$ alloy. **b**, Brillouin zone for CuPt alloy with space group $R\bar{3}m$. **c**, Bulk band dispersions of CuPt-order PbSnTe$_2$ alloy under 1.13% tension strain are shown along high-symmetry lines $\bar{Z}$ (0 0 0.5)- $\bar{\Gamma}$(0 0 0)- $\bar{F}$(0 0.5 0.5)- $\bar{L}$ (0 0.5 0)- $\bar{K}$ (0.31,0.66,0.66) calculated by the modified Becke and Johnson exchange potential. In the CuPt structure, the four equivalent L points in the rocksalt phase project into one $\bar{\Gamma}$ and three equivalent $\bar{F}$ points (i.e. "4=1+3"). The bottom plots show the band character of conduction band edge and valence band edge near the $\bar{\Gamma}$ and $\bar{F}$ points, respectively. The wide green and narrow red lines indicate the band is dominated by the anion (Te) $p$ and cation (Sn and Pb) $p$ states, respectively, and corresponding parities are also labeled. **d**, **e**, **f** and **g**, The charge density distribution of the valence band edge (**d, f**) and conduction band edge (**e, g**) at the $\bar{\Gamma}$, and $\bar{F}$ point, respectively. The value of the isosurfaces is set to $3\times10^{-3}$ e/Å$^3$.

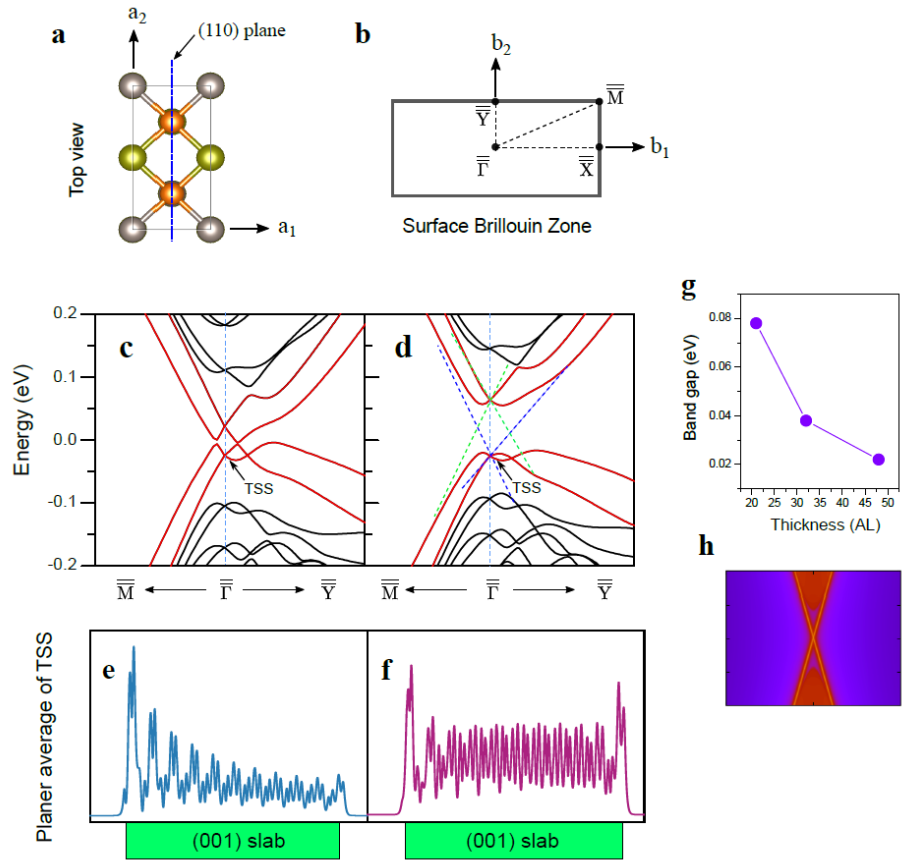

FIG. 3. **Topological surface band structures of a 21-atomic-layer thick slab of the CuPt-ordered PbSnTe2 alloy in the TCI and TI phases, respectively. a**, Crystal structure (top view) of the 21-atomic-lyaer thick slab made of the CuPt-ordered PbSnTe$_2$ alloy along the [001] direction. A blue dashed line indicates the (110) mirror plane.. **b**, The 2D surface BZ of the slab. A double-overline is used to mark the high symmetry k-points in the surface BZ, distinguishing from ones in reduced 3D alloy BZ (marked with single overline) and ones in the parent rocksalt BZ. **c,** The topological surface band structure of the slab in the TCI phase. **d,** The topological surface band structure of the slab in the TI phase. **e, f,** The topological surface states located at zone center for TCI and TI phases as indicated by arrows in c and d, respectively. **g,** The magnitude of the topological bandgap in the $Z_2$ phase as a function of slab thickness. **h,** $k \cdot p$ Hamiltonian model predicted Dirac cone on the surface of the semi-infinite $PbSnTe_2$ alloy in the TI phase.